\documentclass[apj,numberedappendix]{emulateapj}
\usepackage{natbib}
\usepackage{amsmath}
\usepackage{multirow}

\usepackage{bm}
\usepackage{color}
\usepackage{dcolumn}
\usepackage{graphicx}
\usepackage{hyperref}

\newif\ifAMStwofonts
\AMStwofontstrue

\def\vnabla{\bm{\nabla}}
\def\vD{\bm{\Delta}}

\def\vB{{\bf B}}

\def\gsim{~\rlap{$>$}{\lower 1.0ex\hbox{$\sim$}}}

\def\tl{\tilde l}

\def\simpropto{\lower.2ex\hbox{$\; \buildrel \propto \over \sim \;$}}
\def\ltsim{\lower.5ex\hbox{$\; \buildrel < \over \sim \;$}}
\def\gtsim{\lower.5ex\hbox{$\; \buildrel > \over \sim \;$}}
\def\ltsim{\lower.5ex\hbox{$\; \buildrel < \over \sim \;$}}
\def\gtsim{\lower.5ex\hbox{$\; \buildrel > \over \sim \;$}}

\def\vnabla{{\bf \nabla}}

\def\kms{\mbox{km\,s$^{-1}$}}

\def\dd{\,{\rm d}}

\def\kms{\ {\rm km\,s^{-1}}}
\def\hmpc{\ {\rm h^{-1}Mpc}}

\def\tl{ {\tilde l}}

\def\dd{{\rm d}}

\def\grad{\vnabla}
\def\pmb#1{\setbox0=\hbox{#1}%
\kern-.025em\copy0\kern-\wd0
\kern.05em\copy0\kern-\wd0
\kern-.025em\raise.0433em\box0}

\def\vv{\pmb{$v$}}
\def\vA{\pmb{$A$}}
\def\vv{\pmb{$v$}}

\def\vs{\pmb{$s$}}

\def\vx{\pmb{$x$}}
\def\vy{\pmb{$y$}}
\def\vz{\pmb{$z$}}
\def\vr{\pmb{$r$}}

\def\vS{\pmb{$S$}}

\def\hvr{\hat {\vr}}
\def\hvs{\hat {\vs}}

\def\hvx{\hat {\vx}}
\def\hvy{\hat {\vy}}
\def\hvz{\hat {\vz}}

\def\simlt{\lower.5ex\hbox{$\; \buildrel < \over \sim \;$}}
\def\simgt{\lower.5ex\hbox{$\; \buildrel > \over \sim \;$}}

\def\vnabla{\pmb{$\nabla$}}

\newcommand{\beq}{\begin{equation}}
\newcommand{\eeq}{\end{equation}}
\def\beqa{\begin{eqnarray}}
\def\eeqa{\end{eqnarray}}
\def\fixit#1{}
\def\hmpc{h^{-1}\,{\rm Mpc}}

\def\dd{{\rm d}}

\shorttitle{ LG motion  from redshift surveys }
\shortauthors{Nusser, Davis \& Branchini}
\begin{document}
\title{On the recovery of Local Group  motion  from galaxy redshift surveys}
\author{Adi Nusser }
\email{adi@physics.technion.ac.il}
\affil{Physics Department and the Asher Space Science Institute-Technion, Haifa 32000, Israel\\
e-mail: adi@physics.technion.ac.il}
\author{Marc Davis}
\email{mdavis@berkeley.edu}
\affiliation{Departments of Astronomy \& Physics, University of California, Berkeley, CA. 94720}
\author{Enzo Branchini}
\email{branchin@fis.uniroma3.it}
\affiliation{Department of Physics, Universit\`a Roma Tre, Via della Vasca Navale 84, 00146, Rome, Italy\\
INFN Sezione di Roma Tre, Via della Vasca Navale 84, 00146, Rome, Italy\\
INAF, Osservatorio Astronomico di Roma, Monte Porzio Catone, Italy}


    
\begin{abstract}
There is a $\sim 150\kms$ discrepancy between the measured  motion of the Local Group of galaxies (LG) with respect to the CMB
and the linear theory prediction  based on the gravitational force field of the large scale structure in full-sky redshift surveys. 
We perform a variety of tests which show that  the LG  motion  cannot be recovered to better than 
$150-200\kms$ in amplitude and within a $\approx10^\circ$ in direction. 
The tests rely on catalogs of mock galaxies identified in the Millennium simulation using semi-analytic 
galaxy formation models. We compare these results to  
 the $K_s=11.75$ Two-Mass Galaxy Redshift Survey, which provides the deepest, widest and most complete
spatial distribution of galaxies available so far.
In our analysis we use a new, concise relation for deriving the LG motion  and 
bulk flow from the true distribution of galaxies in redshift space. 

Our results show that the  main source of uncertainty is the small effective depth of  surveys like the 2MRS that
prevents  a proper sampling of the large scale structure beyond $\sim100\hmpc$. Deeper redshift surveys are needed to 
reach the  ``convergence scale" of  $\approx 250\hmpc$ in a $\Lambda$CDM universe.
Deeper survey would also mitigate the impact of the ``Kaiser  rocket" which, in a survey like 2MRS,
remains a significant source of uncertainty.
Thanks to the quiet and  moderate density environment of the LG, purely dynamical uncertainties of the linear predictions
are subdominant at the level of $\sim 90\kms$. 
Finally, we show that deviations from linear galaxy biasing and shot noise errors provide a minor contribution to the total error budget.

 \end{abstract}

\keywords{Cosmology: theory, observations, large scale structure of the universe, dark matter}
\maketitle
\section{Introduction}   
\label{sec:int}

The group of galaxies containing M31, the Milky Way (MW) and about a dozen other, 
much smaller,  galaxies (excluding satellites) within $\sim 1.4$ Mpc 
form a bound system which is detached from the general cosmic expansion \citep{yst77}.
This Local Group (LG) of galaxies resides in mildly over-dense region characterized by a
remarkably small velocity shear. 
Just like any other cosmological object, the LG  is expected to move with a non-vanishing velocity relative to the 
general expanding background. 
The best approximation to the frame of reference defined by the cosmological background is undoubtedly based on 
temperature maps of the Cosmic Microwave Background (CMB).
The high degree of dipole anisotropy in the temperature of the CMB on the sky
\citep{1993ApJ...419....1K, 1996ApJ...473..576F, 2009ApJS..180..225H} is interpreted as a Doppler boosting resulting from our motion through a  highly isotropic thermal CMB photons. 
This interpretation  has recently been reinforced  by the detection of
the corresponding   modulation and aberration of the CMB fluctuations
 observed by the Planck satellite \citep{2013arXiv1303.5087P}.  The observed dipole yields a very precise measurement of the Solar barycenter velocity relative to  the ``CMB frame" of reference in which an observer at rest would measure a vanishing  dipole anisotropy. 
Augmented with astronomical estimation of the LG motion relative to the Sun (c.f. \S\ref{sec:refs}), the CMB dipole 
provides  $V_{\rm lg}=627\pm 22\kms$ toward $(l,b)=(276^\circ\pm 3^\circ,30^\circ\pm 3^\circ)$ as the reference value for 
 the LG motion relative to the CMB frame \citep{1993ApJ...419....1K}. In the standard cosmological paradigm \citep{Peeb80}, the LG is accelerated by   
the cumulative gravitational pull of the surrounding large scale structure. In linear theory, the peculiar velocity is
proportional to the peculiar gravitational force field times the Hubble time, with proportionality factors depending on 
the background  mass density.
Therefore, it is natural to ask whether the observed large scale structure, as traced by the galaxy distribution, 
could indeed account for the LG motion.

This issue was recently studied by \cite{DN10} who found good agreement between the local velocity and gravitational fields, in contrast to the earlier data with an inferior velocity field, which gave 
irreconcilable differences between the two (\cite{DNW96}).
The earlier studies of  \citep{1980ApJ...242..448Y,DH82} 
 addressed the gravity versus velocity fields, with somewhat contradictory results, using either the 
angular positions and fluxes from photometric galaxy catalogs 
\citep{yahil86,MD86,harmon87,Vill87,P88,bilicki11}
or the full 3D distribution of different types of extragalactic objects
ranging from the infra-red selected galaxies of the IRAS 1.2 Jy catalog
\citep{strauss92,web97,zaroubi99} 
and extension to fainter fluxes, the PSC$z$ catalog 
\citep{schmoldt99,schmoldt99b,rr00,bp06}, 
optically selected galaxies
\citep{Lahav87,Lybell89,huds93}, 
mixed catalogs of infra-red and optical galaxies 
\citep{Lahav88,dmellow04},
galaxy clusters selected from optical plates 
\citep{PV91,scaramella91,scaramella94,bps96,dale99} 
and, finally, X-ray selected galaxy clusters
\citep{pk98,ke06}. 

%
 
In this work we focus on determining how well the LG motion 
 can be recovered from  the observed galaxy distribution via linear instability theory.
 Detection of significant departures from theoretical expectations require a complete understanding and characterization of all possible error sources.
In the case of cosmological dipoles these are \citep{schmoldt99}
\begin{itemize}
\item Cosmic variance from finite volume sampling.
All sky galaxy surveys become significantly more dilute at larger distances, limiting the depth within which density fluctuations can 
be reliably probed for the reconstruction of the LG motion.   
\item Shot noise from sparse sampling of the mass tracers.
\item Highly non-linear dynamical effects that spoil the relation the tight relation between the gravity and the LG motion.
 They include nonlinear motions and growth of density fluctuations.
 \item Deviations from strict linear biasing between galaxies and the underlying mass density fluctuations. 
\item Observational uncertainties and biases arising from incompleteness and selection effects in the parent objects' catalogue.

\end{itemize}
 
 The most appropriate  route to estimate the impact of these uncertainties and  assess the adequacy of linear theory
has   been pioneered by \cite{1991ApJ...372..394D}. It
relies on  the extensive use of realistic mock galaxy catalogs extracted from N-body simulation, 
since they simultaneously account for non nonlinear effects as well as selection effects specific to 
the specific dataset.
In particular, the mocks should have LG candidates residing in a  mild density region  and small velocity shear, as in the real data.
Although,  one can attempt to account for  non-linear dynamical effects within a likelihood formalism by quantifying the "decoherence" between 
the gravity and the peculiar velocity fields \citep{chodor02,chodor04,chodor08}, we opt here to rely on mock catalogs
extracted from fully nonlinear $N$-body simulation.
While we aim at a general discussion of the problem, we shall consider here the case of
 the  Two-Mass  Redshift Survey  (2MRS) \citep{huchra12},  i.e. the 
 the deepest nearly-all sky survey of angular positions and spectroscopic galaxy redshifts limited to $K_s=11.75$
 and arguably the best sample of objects to estimate the LG motion.

The outline of the paper is as follows. 
A brief description of how the  LG motion has been measured in the literature is given in \S\ref{sec:refs}.
In \S\ref{sec:equations}, we review the linear theory predictions and
offer a new useful relation for deriving the LG motion from a given distribution of mass tracers in redshift space. 
In \S\ref{sec:mocks}  we first test  linear theory  as measured by the full dark matter distribution  and the full volume-limited galaxy distribution.
We characterize the biasing relation of the  galaxy catalogs, address the reliability of linear reconstruction of the LG motion,  and assess the impact 
shot noise errors. In \S\ref{sec:dmrec}  we consider the analysis when applied to  mock 2RMS  catalogs.
The results are presented in  \S\ref{sec:results} in which we compare the LG motions obtained from
the distribution of mock galaxies to that measured directly in the N-body simulation.
The impact of the so-called Kaiser effect is outlined in a separate \S\ref{sec:kaiser} and, finally,
we end with a general discussion in \S\ref{sec:disc}.
In the Appendix the reader will find a detailed derivations of the linear theory relations used here and farther discuss the 
dependence on the so-called distortion parameter $\beta$ of peculiar velocities reconstructed from redshift space data.

\section{The Local Group}
\label{sec:refs}
The identification of the LG of galaxies and determining its motion in the heliocentric and CMB frames have 
 been the subject of research 
 for many years
\citep[e.g.][]{1956AJ.....61...97H,yst77,1986ApJ...307....1S,1996AJ....111..794K,1998MNRAS.298..114R,1999AJ....118..337C,2000CAS....35.....V}.
Obvious galaxy members of the LG are M31 and the MW. 
The assignment of other members  to the LG is  based on the assumption that they form
a bound object detached from the general cosmic expansion.  
Therefore, the LG motion is found by fitting a constant velocity to the radial velocities (redshifts) of LG galaxies measured  with respect to the Local Standard of Rest (defined by mean motion of stars in the solar neighborhood). 
The criteria of whether or not a galaxy belongs to the LG depends on the LG velocity itself. Hence the procedure is 
essentially iterative. 
An LG member galaxy  must satisfy  the following criteria \citep[e.g.][]{yst77,1981Obs...101..111L,1999AJ....118..337C}: 
\begin{itemize}
\item it should not appear to be associated with any other group of galaxies;
\item  its distance from the LG barycenter should be smaller than the radius of the surface of zero velocity (with respect to the LG velocity). 
\item  its radial velocity does not deviate significantly from the value obtained from the constant velocity fit. 
\end{itemize}
A recent analysis (Mikulizky \& Nusser 2013)  yields 14 well-established members within $\sim 1.4$ Mpc from the LG barycenter, not including satellites of M31 and MW. The velocity of LG in the CMB frame from this analysis is  $622\pm 33 \kms$ in the 
$l=277^\circ\pm 3^\circ$ and   $b=33^\circ\pm 3^\circ$ direction, in agreement with \cite{1993ApJ...419....1K}.
In the standard paradigm for structure formation, this motion should be the result of the cumulative gravitational tug of 
matter in the universe.
The subject of the paper is to assess how well the gravitational force field matches the LG velocity measured in the  CMB.

\section{Linear theory of the LG motion}
\label{sec:equations}

We assume that all quantities are given at the present time with the expansion  scale factor $a$ set 
to unity so that the comoving and physical distances are equal. 
Let $\vr$ be the real space coordinate (proper distance) and $\vv=\dd \vr /\dd t$  the corresponding peculiar
velocity of a patch of matter. We shall assume that the mass  density  contrast, $\delta_{dm}$, is related to 
the galaxy  (number) density contrast, $\delta$,   by a linear biasing relation 
  $\delta=b\delta_{dm}$. 
The linear theory  for structure formation \citep[e.g.][]{Peeb80} relates the divergence of the peculiar velocity field, $\vv(\vr)$, to   
the density contrast as
\begin{equation}
\label{eq:linx}
\vnabla \cdot \vv =-H_0 \beta\delta\; ,
\end{equation} 
where $H_0$ is the Hubble constant, $\beta=f/b$ and $f=\dd {\rm ln} D/\dd {\rm ln }a$ is the logarithmic derivative 
of the linear growth rate with respect to  the scale factor, $a$. 

Observations provide the angular positions and redshifts of  galaxies,
 $cz=H_0 r+\hvr \cdot \vv$ where $\hvr\cdot \vv$ is  the radial peculiar velocity, and $\hvr$ denotes the direction of vector $r$.
Hence, for the realistic reconstruction of velocities from redshift space data, 
the relation (\ref{eq:linx}) needs to be modified in order to account for the added displacement  from $r $ to $cz$.
We define the redshift space coordinate, 
\begin{equation}
\vs \equiv \vr+{H_0}^{-1}({\hvs\cdot \vv}) \hvs\; 
\label{eq:xtos}
\end{equation}
where $\hvs=\hvr$. 

Let  $n^s(\vs)=\bar n [1+\delta^{\rm s}(\vs)]$ and $n(\vr)=\bar n [1+\delta(\vr)]$ be, 
respectively, estimates of the number densities  of galaxies in redshift  and  real space, where $\bar n$ is 
the mean number density of galaxies in the survey (assumed to be the same in both spaces). 
In the limit  $\delta \ll 1$,  the  mapping (\ref{eq:xtos}) and the  
 continuity equation  $n^s(\vs)\dd^3 s=n(\vr)\dd^3r$ modifies the 
  real space linear equation (\ref{eq:linx}) as
\begin{equation}
\label{eq:lins}
\vnabla \cdot \vv +\beta\vnabla \cdot  [(\hvs\cdot \vv) \hvs]=-H_0 \beta \delta^{\rm s}\; .
\end{equation}
To linear order, $\delta^{\rm s}(\vs)=
\delta^{\rm s}(\vr)$ and similarly for $\delta$ and $\vv$. 
Given appropriate boundary conditions, a unique solution to this equation can be obtained 
 for a potential flow,  $\vv=-\grad \phi$, where $\phi$ is a scalar function 
of the spatial coordinates.
For our purposes, it is convenient to express the solution in terms of the spherical harmonics, $Y_{lm}$,  expansion of the angular dependence 
of $\phi(\vr)$ and $\delta(\vr)$. Writing
\begin{equation}
\phi(\vr) \equiv \phi(r,\hvr)=\sum_{{l}\ge 0}\sum_{{\rm m}=-{l}}^{l}\phi_{lm}(r)Y^*_{lm}(\hvr) \; ,
\nonumber
\end{equation}
where $\hvr$ is the radial unit vector,  $r=|\vr|$ the distance. 
$\delta(r,\hvr)$ is similarly expanded.
The solution to (\ref{eq:lins})  is \citep{ND94} 
\begin{eqnarray}
\nonumber  \phi_{lm}(s)&=& - \frac{H_0\beta/(1+\beta)}{2\tl+1}
\left[   s^\tl\int_{s}^{s_1} \frac{\delta^{\rm s}_{lm}(u)}{u^{\tl-1}}\dd u 
 \right.  \\
 &+&   \left.  \frac{1}{s^{\tl+1}} \int_0^s \delta^{\rm s}_{lm}(u)u^{\tl+2}\dd u     \right]\; ,
 \label{eq:vlins}
 \end{eqnarray}
 where $\tl\le l$ is related to the harmonic order $l$  through the algebraic equation 
 $(1+\beta)\tl(\tl+1)-{l}({l}+1)=0$, and $s_1$ is a constant dictated by the appropriate boundary conditions.
The solution in real space is  obtained   in the limit $\beta \ll 1$ where 
$\tl \rightarrow  l$ and $\beta/(1+\beta)\rightarrow  \beta$.
Only the dipole, $l=1$, component is relevant for the LG motion.  For this component, the appropriate choice is  $ s_1=0$. Thus we  work in the freely falling LG frame, not the CMB frame, because obtaining the CMB should be a {\it result} of the analysis.  After all, we don't know how large is the sphere around us that has the same dipole CMB anisotropy. Further, working with CMB redshifts causes a singular behavior of the density at the origin. 
In the LG frame, the mean motion of a very distant thin spherical shell, i.e. the {\it reflex dipole},
approaches the negative of the LG motion in the CMB frame. Thus the LG motion in the CMB frame, $\vv_{\rm lg}$, is identified with the negative of 
the reflex dipole of a very distant shells. 
In \S\ref{sec:appendixa} we derive a new relation for relating $\vv_{\rm lg}$ 
to the density distribution in redshift space. 

For a sampling of the density in redshift space 
by   a discrete distribution of $N$ mass tracers (galaxies) with redshift coordinates $\vs_i$ ($i=1\cdots N$) the relation yields
\begin{eqnarray}
\vv_{\rm lg}(R_{\rm out})&=& \frac{R_{\rm out}^{\tl-1}}{(1+\beta)}\frac{\beta H_0}{4\pi\bar n}\sum_{R_{\rm lg}<s_i<R_{\rm out}}\frac{\vs_i}{\varphi_i s_i^{\tl+2}}\\
 \nonumber  &-&\frac{(1 -\tl)R_{\rm out}^{-(\tl+2)}}{(1+\beta)(2\tl+1)}\frac{\beta H_0}{4\pi\bar n }\sum_{R_{\rm lg}<s_i<R_{\rm out}}\frac{s_i^{\tl-1}}{\varphi_i}\vs_i \; ,
\label{eq:vlgs} 
\end{eqnarray}
where $\tl$ correspond to $l=1$, i.e. it is the solution to $(1+\beta)\tl(\tl+1)-2=0$, the selection function
$\varphi_i$ compensates for the missing faint galaxies in flux limited surveys and $\bar n$ is a measure of the average number density of galaxies. The sum extends over all tracers between $R_{\rm lg}$, the radius assigned to the LG, and 
a maximum distance $R_{\rm out}$.
In principle all fluctuations  out to  $R_{\rm out}\rightarrow \infty $  contribute to $\vv_{\rm lg}$. In a hierarchical model for structure formation, distant structures typically make smaller contributions. 
It is therefore, interesting to study the convergence of $\vv_{\rm lg}$ as a function of $R_{\rm out}$. 
Further, in realistic redshift surveys, 
the noise increase dramatically at large redshifts, due the significant decrease in the number of observed galaxies
and the recovery of $\vv_{\rm lg}$  can be assessed reliably  only out to  the  radius beyond which the large scale structure is poorly probed.

The real space counterpart of the relation (\ref{eq:vlgs}) is  obtained by 
setting $\tl=l=1$ and replacing $f$ with  $\beta/(1+\beta)$. This yields 
\begin{equation}
\vv_{\rm lg}=\frac{H_0\beta}{4\pi \bar n} \sum_{R_{\rm out}>r_i>R_{\rm lg}}\frac{\vr_i}{\varphi_i r_i^3}\; ,
\label{eq:vlg}
\end{equation}
where we use the same symbol $R_{\rm out}$ to indicate the maximum distance in real space.
Note  the disappearance of the counterpart of the second term on the r.h.s of equation (\ref{eq:vlgs}).
However, even for redshift space reconstruction by equation (\ref{eq:vlgs}), the second term makes  
negligible contribution to $\vv_{\rm lg}$  as we find in the numerical tests below.


\section{Construction of Mock galaxy catalogs}
  \label{sec:mocks}

 
We consider mock catalogs designed to match the  
2MRS catalog of $\sim 45000$ galaxies
with $K_s\le 11.75$ \citep{huchra12}. 
A parent simulated catalog of the whole 2MASS catalog has been
prepared \citep{delucia} by incorporating semi-analytic galaxy formation models
in the Millennium simulation \citep{mill} of  the $\Lambda$CDM model with  $\Omega=0.25$, $\Omega_b=0.045$, $\sigma_8=0.9$,  $\Lambda=0.75$, and $H_0=73\kms {\rm Mpc}^{-1}$.
From this parent catalog we have drawn 53 independent mock 2MRS  catalogues 
satisfying the following conditions:
\begin{itemize}
\item The ``observer" in each mock is selected to reside in a galaxy with 
a quiet velocity field within $500\kms$,  similar to the observed universe.  
One observational signature of this quiet flow is that in the LG  frame, the only
 galaxies with measured negative redshift belong to the Virgo cluster.
We enforce this condition by imposing that 
 the observer  sees at most one cluster 
 that has high enough peculiar velocities to result in negative redshifts.   
\item  Note that in finding a cluster which produces negative redshifts in its core region necessarily implies the average overdensity toward the region is substantial. For example,  the average overdensity towards the Virgo cluster is $\delta \sim 2$  \citep{DH82}, 
and the mock catalogs are  roughly the same.  That means that the flows in this direction are becoming nonlinear.
To keep the overdensity close to that of the Virgo cluster, we select only clusters with mean overdensity between the LG and the cluster center to be $\delta < 2$. This is quite a stringent constraint that, alone, eliminates $\sim 70$ \% of the potential LG "observers".
\item The velocity of the central galaxy must be in the range 400 to 700 $\kms$ to match that of the LG with respect to the CMB.
\item The density in the LG environment, i.e. averaged over a sphere of 5 Mpc radius around the observer,  is  less than twice the cosmic mean. 
\item We impose the  constraint that  the bulk flow of a sphere of radius, $3.5\hmpc\approx 5\rm Mpc$ ($h=H_0/[100\kms {\rm Mpc}] \approx 0.7$), centered on the 
 LG, is also in the range $400-700\kms$. This radius is more than three times larger than the radius assigned to the real LG ($\sim 1.4\rm Mpc$). This additional constraint is meant  to eliminate strong nonlinearities that may still be present in some of the mocks after applying  the previous  constraints mentioned above. 
This choice is justified by the fact that strong nonlinearities inside $5$ Mpc seem to be absent in  the real Universe,
as indicated by the fact that the flow is fairly quiet within that radius. 
Hereafter we take $R_{\rm lg}=5\rm Mpc$  as the radius of the LG and treat the motion of 
of the central sphere of that  radius as the motion of the LG. 
\end{itemize}
 
The mocks are taken from the $z=0$ simulation output and, therefore, are  free from any possible galaxy evolution.   The 2-point correlation function of the mock galaxies fits reasonably well the observed one \citep{w09}, but less so is the K-band luminosity function, resulting in a discrepancy with the observed number of galaxies.
To fix the problem, the original luminosity of mock galaxies was shifted to brighter values by  $\sim 1.5$ magnitude. 
We obtained, on average,  $\sim 50000$
galaxies per mock, slightly larger than but close to the  real $K_s=11.75$ 2MRS catalog. 
Each  of the 53 mock catalogs contains galaxy distances, peculiar velocities (and hence redshifts), angular positions and $K_s $-band magnitudes. 

\subsection{The selection function}
\label{sec:self}

In the application  to a flux limited survey like 2MRS, each galaxy in the summation in the relations (\ref{eq:vlg}) and (\ref{eq:vlgs}) should be weighted by 
the inverse of the selection function, $\varphi$, to compensate for missing faint galaxies that fall below the flux limit. 
The selection function depends on the  galaxy distances and it is physically determined by the distribution of   galaxy luminosities. 
In the mocks, where  galaxy distances and apparent magnitude are both known, 
we compute $\varphi$ using  a direct method which avoids the explicit calculation of the luminosity 
function \citep{turner79,KOS,DH82}. The method provides  discrete values of $\varphi$ in distance bins, which are then interpolated to 
the galaxy distances to yield the weights to be assigned to individual galaxies. 
In realistic applications, however, the distances to the galaxies are not known.
Using redshifts rather than distances as arguments to the selection function induces systematic errors, sometimes dubbed as "Kaiser Rocket" effect \citep{K87}.  The hazards of not accounting explicitly  for the Kaiser rocket effect are given in \S\ref{sec:kaiser}.

\subsection{Bias of the selected galaxies} 
\label{sec:biasing}

Galaxies typically  form  
at the peaks of the mass density field and, therefore, are not  unbiased tracers of the underlying density field. 
An indirect but strong  evidence for galaxy biasing is the  fact that different types of object exhibit different clustering properties.
On large scales, however, it is safe to assume a linear biasing relation between the density contrast of 
the matter and the galaxy distribution,  $\delta({\rm galaxies})=b \delta({\rm mass})$, with a constant bias factor $b$.
If biasing is a local, though not necessarily a Poisson, process, then this form is motivated by theory on linear scales
 \citep[e.g.][]{K88,1993MNRAS.262.1065C,1993ApJ...413..447F,1998ApJ...504..607S,1999ApJ...521L...5C,2001MNRAS.325.1359S,2007PhRvD..75f3512S},
  confirmed by numerical experiments \citep[e.g.][]{kn97,2000ApJ...528....1N,2000MNRAS.311..793B,2007APh....26..351H} and supported by observations involving galaxy samples 
dominated, like in the 2MRS case, by late type galaxies \citep[e.g.][]{2001PhRvD..63d3007T,2002MNRAS.335..432V,w09}.

\begin{figure} 
\includegraphics[width=0.48\textwidth]{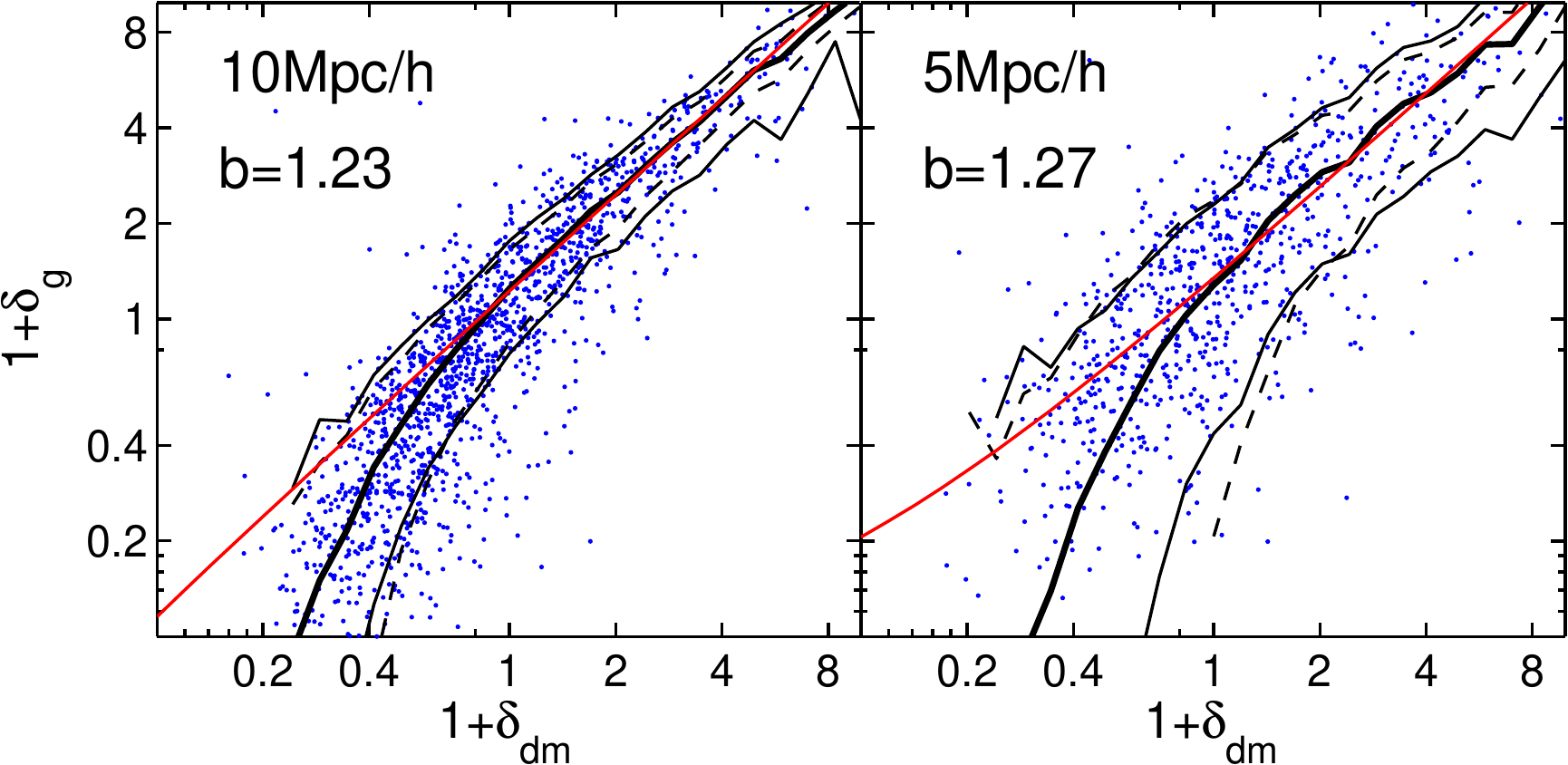}
\caption{A scatter plot (logarithmic scale) of the galaxy  versus the dark matter over-densities
in  the simulation. For each  of the 53 mocks, densities of 125  randomly selected points with a distance  
$<70\hmpc$ from the LG candidate, are shown. The left and right panels correspond to  densities in cubic cells 
of $10\hmpc$ and $5\hmpc$ on the side, respectively.
 The thick solid curve in each panel is the mean of $1+\delta_{\rm g}$ 
at a given $1+\delta_{\rm dm}$. The two thin solid curves are  $\pm 1\sigma $ scatter computed from points above and 
below the mean.  Dashed curves are the expected $\pm 1\sigma$ Poisson (shot-noise) scatter.
The nearly straight red lines show $\delta_{\rm g}=b\delta_{\rm dm}+const$, where $b$ (indicated in the figure) are determined using linear regression  from 
 points in the range $-0.5<\delta_{\rm dm}<4$.  
    }
 \label{fig:bias}
\end{figure}

We explore here  the bias of the distribution of the  mock galaxies with
respect to the dark matter density field in the simulation.
Gerard Lemson has kindly 
used the facilities of the Millennium Simulation Database
to produce for us the density field from  all  $2160^3 $ dark matter particles in the simulation box
on a cubic grid of $1\hmpc$ spacing.  
Density fields from the distribution of mock galaxies have also been computed directly 
for all the mocks. Figure (\ref{fig:bias})  is a scatter plot of the over densities computed from the mock galaxy distribution versus the dark meter density field. 
For $\delta_{\rm dm}\ltsim 3$, the scatter in the relation is mainly  Poissonian. However, at higher densities, intrinsic 
scatter in the biasing relation dominates. 
The relation in small (right panel) and large (left) cells is fairly linear, $\delta_{\rm g}=b\delta_{dm}$,  in the moderate density ($-0.2\ltsim \delta_{\rm dm}\ltsim 4$ in the two panels) regions, 
 with a weak dependence of $b$  scale: $b\sim 1.23$ and $ 1.27$ in the large and small cells, respectively. 
 The values change according to the density cut used in fits. Exploration of $b$ for various densities yields $
 1.2<b<1.35$ as an acceptable range.
 We shall continue to  assume  linear bias, bearing in mind that it breaks down in deep voids.  

\section{Reconstruction of the LG motion}
\label{sec:recons}

In this section we tests the ability  to reconstruct the LG velocity using the relations  (\ref{eq:vlgs}-\ref{eq:vlg}). We first consider the 
ideal case in which we can perform the reconstruction using the  dark matter density field. The aim of this test is to assess the 
 possibility of reconstructing  $\vv_{\rm lg} $ and to estimate the impact of shot noise errors. 
 In the second part we repeat the reconstruction procedure on the realistic 2MRS mocks described in Section~\ref{sec:mocks}.
The goal here is twofold:  to assess our ability to determine $\beta$ by comparing the predicted LG motion with the true LG velocity and 
 to measure the accuracy with which one can predict  $\vv_{\rm lg} $ when the $\beta$  is given {\it a priori}.

\subsection{Reconstruction tests using the full dark matter distribution}
\label{sec:dmrec}

 Given the location of each mock  LG in the parent simulation we use 
 the actual density field $\delta_{\rm dm}$ to recover the motion of the corresponding LG. 
 We perform this test only in real space, by adapting  
equation  (\ref{eq:vlg}) to density fields given on a grid, i.e.

\begin{equation}
\label{eq:vlgdm}
\vv_{\rm lg}(R_{\rm out})=\frac{H_0 \beta}{4\pi} \sum_{R_{\rm out}>R_{\alpha}>R_{\rm lg}}  \delta_\alpha \frac{\vr_\alpha}{r_\alpha^3}
\end{equation}
where the summation is over the grid points, $\vr_\alpha$ is the distance of the grid cell $\alpha$ from the LG position and $\delta_\alpha$ is $\delta_{dm}$ in the cell 
$\alpha$.   
We apply the relation  (\ref{eq:vlgdm}) with the largest possible outer radius, namely $R_{\rm out}=250\hmpc$.
Further,  to minimize  the  influence of mass fluctuations beyond $R_{\rm out}$
we measure $\vv_{\rm lg}$ with respect to the bulk flow of the sphere. Because of the missing power on scales $>500\hmpc $ in the simulation, this 
is only  $10-30\kms$ and hence this last step has little effect on our results.

Since we are dealing with the dark matter directly and are assuming a flat $\Lambda$CDM model,
we have  $\beta=f\approx \Omega_m^{0.55}=0.466$ \citep{Lind05} for $\Omega_m=0.25$ of the simulation. 
However, due to 
nonlinear effects, we do not expect  the reconstructed $\vv_{\rm lg}$ to coincide exactly with its true value.
 We obtain an estimate of $\beta$ in each mock by matching the motion recovered with $\beta=f$,  $\vv^{\rm rec}_{\rm lg}$, 
 with  the true motion, $\vv_{\rm lg}^{\rm tru}$:
 \begin{equation}
 \label{eq:betapar}
  \frac{\beta}{f}=\frac{|\vv_{\rm lg}^{\rm tru} |}{\hat \vv\cdot  \vv^{\rm rec}_{\rm lg}}\; ,
  \end{equation}
Here  $\hat \vv$ is a unit vector in the direction of the true motion and 
   $\hat \vv\cdot  \vv^{\rm rec}_{\rm lg}=v^{\rm rec}_{\rm lg,\parallel}$ is the parallel component of $\vv^{\rm rec}_{\rm lg}$ in the direction of $\vv^{\rm tru}_{\rm lg}$.  
If  the reconstruction error, $\sigma$, scales like $\beta$, as it should, then this estimate of $\beta$ minimizes the quantity
 \begin{eqnarray}
 \nonumber \chi^2(\beta)&=& \sigma^{-2}\left[\vv^{\rm tru}_{\rm lg}-\frac{\beta\vv^{\rm rec}_{\rm lg}}{f}\right]^2\\ 
 \nonumber &=&\sigma^{-2}\left[\left(\frac{\beta \vv^{\rm rec}_{\rm lg,\perp}}{f}\right)^2+\left(v^{\rm tru}_{\rm lg}-\frac{\beta v^{\rm rec}_{\rm lg,\parallel}}{f}\right)^2\right].
 \end{eqnarray}

\begin{figure} 
\includegraphics[width=.48\textwidth]{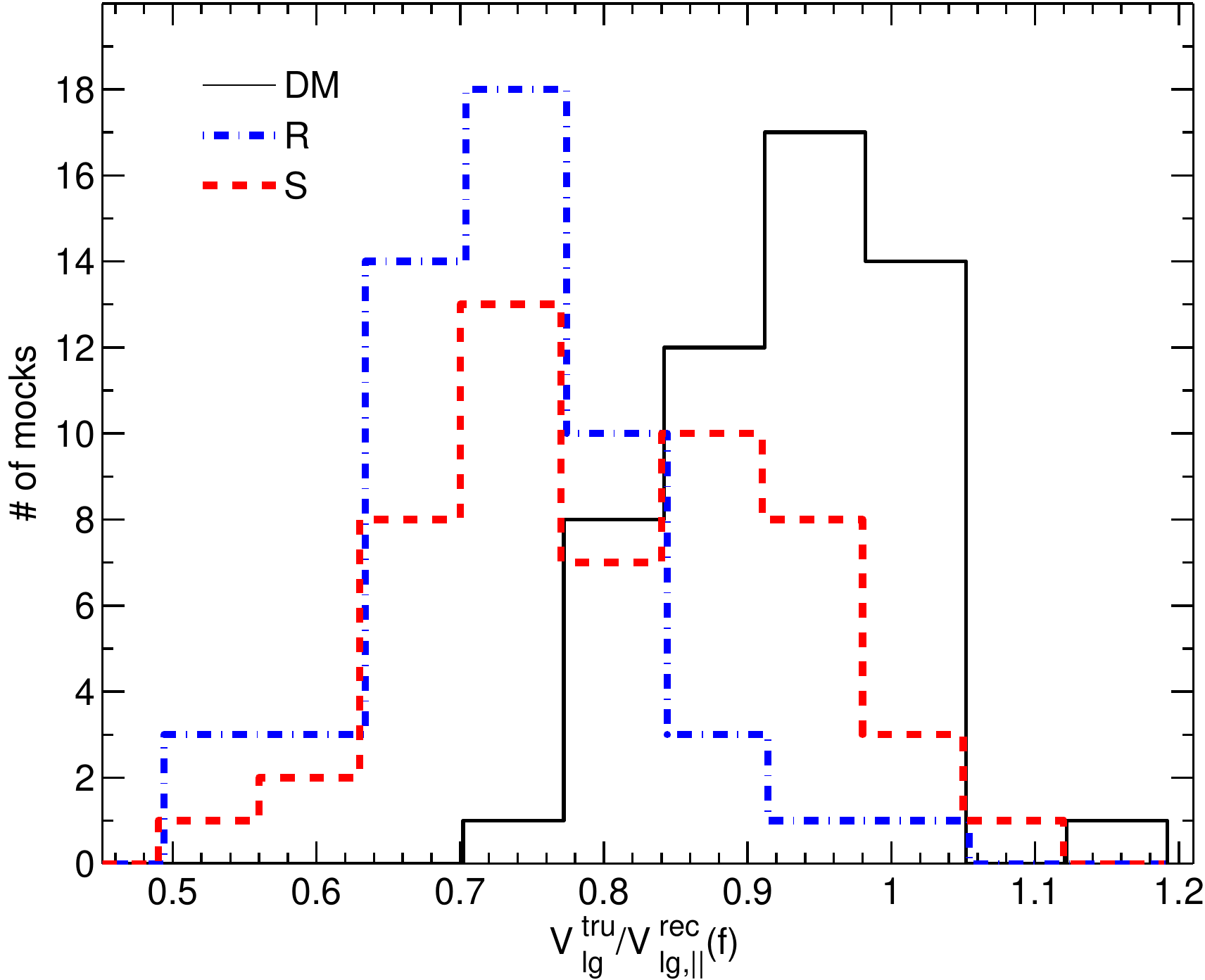}
\caption{Histograms  of the distribution of the ratio $v^{\rm tru}_{\rm lg}$ to the
 $v^{\rm rec}_{\rm lg}$  recovered with a value of 
 $\beta =  f$, i.e. no bias. 
 Black solid, blue dot-dashed  and red dashed curves correspond, respectively, to reconstructions from the  full dark matter density in real space, 
 2MRS mocks in real space, and in redshift space.
 For the real space reconstructions, this ratio equals  $\beta/f$ (see eq.~\ref{eq:betapar}). 
For redshift space, the ratio depends non-linearly on $\beta/f$ as explained in \S\ref{sec:scaling}. 
The averages and standard deviations are (0.72,0.12), (0.80,0.09) and (0.93,0.08), respectively, for the red, blue, and black histograms.
Deviations of the mean values from unity  
quantify the systematic errors  
and random uncertainties in the reconstructions.       
Note that once we account for galaxy bias in the mocks, the mean $\beta$ values  obtained from the histograms are 
consistent with each other. 
 }
\label{fig:beta}
\end{figure}

The solid, black histogram in Figure (\ref{fig:beta}) shows the distribution of $\beta/f$ values  obtained using Equation (\ref{eq:betapar}). The mean value of $\beta/f$ from the 53 mocks is $0.93$ and the $1\sigma$ scatter is 0.085. 
This is a remarkable result considering that the value of  $\beta$ results from a comparison at a single point, i.e. the gravity acceleration at the position of the LG candidate versus velocity of the LG.
The  slight downward bias of $\beta$ with respect to the expected value, $f$,  is due to minor  non-linear dynamical  effects
which persist even in the quiet environment of the LG candidates. 
Since linear theory typically yields larger predicted velocity amplitude \citep{n91}, a smaller value of $\beta$ is obtained 
from the comparison of the  linear prediction with the true velocities.  This deviation from linear theory is less than $10\%$ and would be difficult to detect in any actual test of the LG.
 
\begin{figure} 
\includegraphics[width=0.48\textwidth]{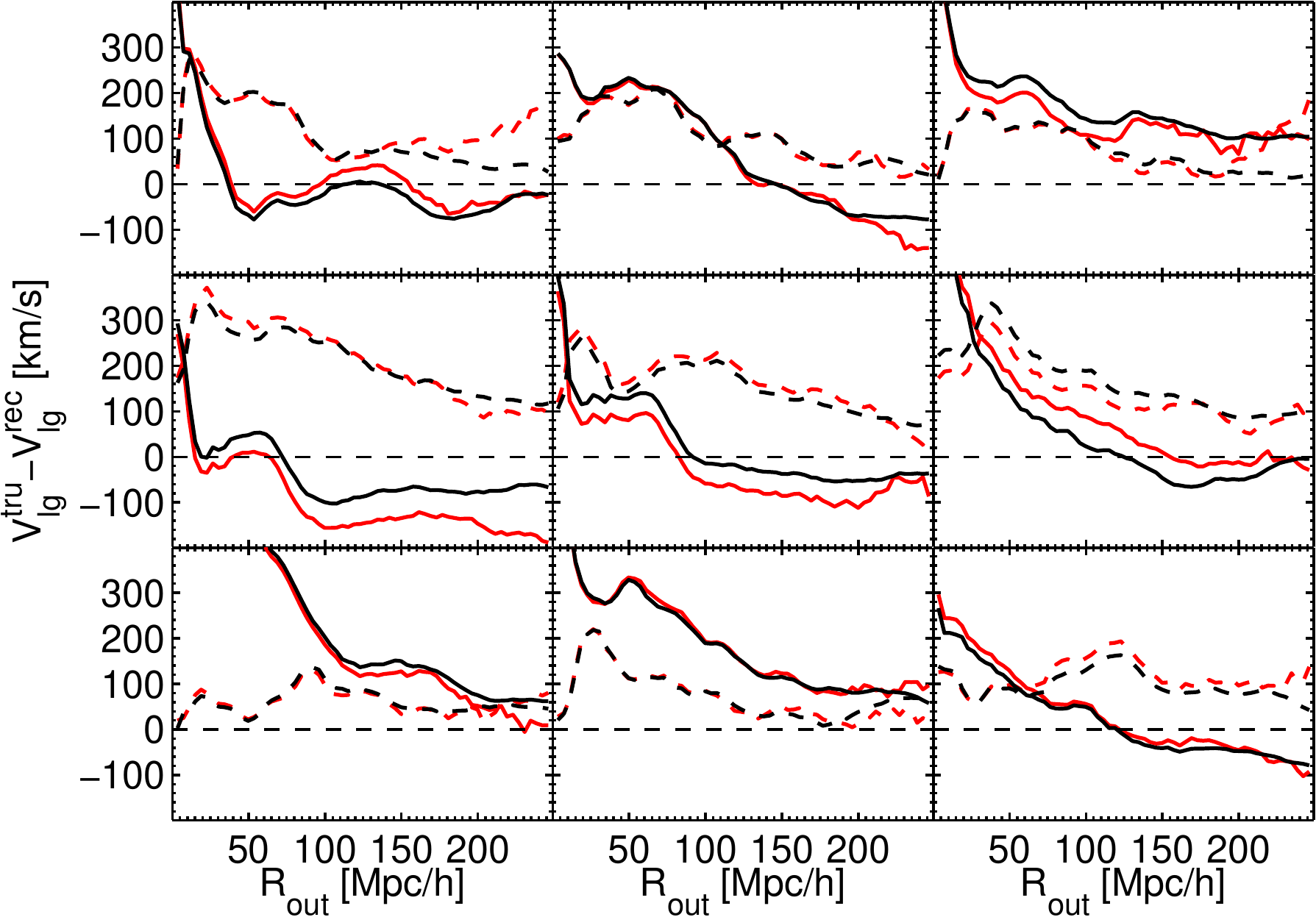}
\caption{LG motions recovered in full density distribution versus dilute distribution of galaxies as a function of $R_{\rm out}$.  
 As $R_{\rm out}$ increases, the velocity residuals should approach zero;  they do shrink but often  asymptote at a value $\sim 100$ km/s.
The  parallel (to the true motion)  and perpendicular components of the residuals are shown as the solid and 
 and dashed  lines, respectively. 
 Black curves correspond to reconstruction from the {\it full} dark matter  density field, while the red curves 
 are obtained by resampling the dark matter with a distance dependent number density of tracers as appropriate
 for the 2MRS.}
\label{fig:vdm}
\end{figure}

The black curves in figure (\ref{fig:vdm}) are the residuals  between true and recovered $\vv_{\rm lg} $ for 9 randomly selected mocks, using the full dark matter density.  In the reconstruction we have adopted $\beta=\beta_{\rm dm }=0.927f=0.43$ for all the mocks. 
The solid curve corresponds to the parallel component, $v^{\rm tru}_{\rm lg,\parallel}-v^{\rm rec}_{\rm lg,\parallel}$ of the residuals 
and the dashed is the amplitude of the perpendicular residual, $\vv^{\rm tru}_{\rm lg,\perp}$.
Both the  parallel and perpendicular residuals change  rapidly for  $R_{\rm out}\ltsim 100\hmpc $ and both flatten 
at $R_{\rm out}\gtsim 150-200\hmpc$.  This shows the danger of working in the CMB frame, as deviations from it are substantial up to $R_{\rm out} \sim 50\hmpc$.

The  residual at $250\hmpc$ is entirely due to dynamical errors in the reconstruction. 
The {\it rms} values of the parallel and perpendicular residuals, at $R_{\rm out}=250\hmpc$, from all 53 mocks are  $54\kms$ and $76\kms$, respectively. The {\it rms} of the total residual, $  \vv^{\rm tru}_{\rm lg}-\vv^{\rm rec}_{\rm lg}$, is $\sigma_{\rm dyn}=94\kms$.

\subsubsection{The impact of shot-noise}
 \label{sec:sn}
 In order to assess the Poissonian (shot-noise)  error introduced by the finite number of galaxies in the mocks, we take each of the 53 mocks and 
 diluted  the dark matter distribution around its LG candidate with a radial distribution that 
 follows the selection function of the corresponding  2MRS mock. 
For each of these dark matter particles-only mocks, we  apply the real space relation (\ref{eq:vlg}) with $\beta=\beta_{\rm dm}=0.43$
to derive $\vv_{\rm lg}$. 
The red curves in  figure (\ref{fig:vdm}) are the residuals  between true and recovered $\vv_{\rm lg}$ obtained from the dilute dark matter particles 
for the same mocks as the black curves. The solid and dashed curves indicate residuals in the
parallel and perpendicular directions, respectively.
The red and the corresponding black curves agree  very  well, indicating  that the shot noise contribute to random uncertainties but does not 
not introduce systematic errors, as expected.
The {\it rms} of the difference  between the reconstructed $\vv_{\rm lg}$ with and without shot-noise 
is  $\sigma_{\rm sn}=90\kms$. This is comparable to  $\sigma_{\rm dyn}=94\kms$ of dynamical inaccuracies in the reconstruction as discussed in the previous subsection. 
Note that the shot-noise error scales linearly with $\beta$ for real space reconstruction. 

Another way to estimate the shot-noise error is by bootstrap resampling of the galaxy distribution in the mocks. This yields the 
following estimate for this error in each mock,  
\[ 
\sigma_{sn}\approx\frac{H_0f}{4\pi \bar n}\left[\sum_{R_{\rm out}>r_i>R_{\rm lg}}\frac{1}{\varphi_i^2 r_i^4} \right]^{1/2}\; .
\]
Both estimates yield similar values.  

The number density of 2MRS galaxies is representative of that in existing and planned spectroscopic galaxy redshift surveys.
A significant reduction of shot noise, say a factor of 2, would require increasing the number density of objects by a factor of 4
which, using the luminosity function in \cite{2012MNRAS.424..472B}, means pushing the magnitude limit of the redshift survey about one magnitude fainter.


\subsection{Reconstruction tests using the realistic 2MRS mock catalogs}
\label{sec:results}

We now turn to the reconstruction of LGs from 
the distribution of synthetic galaxies in the 53 2MRS mocks.
 We perform the reconstruction both in redshift and real space, from
equations  (\ref{eq:vlgs}) and  (\ref{eq:vlg}), respectively. 
Galaxies are  assigned weights according to the selection function 
as outlined in \S\ref{sec:self}.
As before, we remove the effect of external fluctuations beyond the sphere of radius $250\hmpc$ around each mock 
by measuring the LG motion relative to the bulk flow of the sphere. 
Since we want to focus on the ability of linear theory  to recover the LG velocity we shall initially ignore the Kaiser rocket effect
and consider the selection function estimated in real space, deferring
the additional complication related to  the estimation of the selection function in redshift space
to  \S\ref{sec:kaiser}.

\subsubsection{Estimation of  $\beta$ by matching the recovered and true LG motions}

We pursue  the same strategy as in  \S\ref{sec:dmrec} and determine the
values of  $\beta$ by requiring zero residuals in the parallel components of the reconstructed LG velocity
 for each of the 53 mocks. In this way we gauge  
how accurately $\beta$ can be assessed  by matching the gravity field to observed LG motion.    

We reconstruct the LG motion in real and redshift space using Equations (\ref{eq:vlgs}) and (\ref{eq:vlg}), respectively, 
summing over all mock galaxies within  $R_{\rm out}=250\hmpc$ and using $\beta_0=f$ as a reference value.
This  corresponds to assuming that  linear theory applies and that galaxies are unbiased tracers of the mass.
The results of this test are shown  in  Figure~\ref{fig:beta}
in the form of histograms.
They represent the distribution of the ratio $v^{\rm tru}_{\rm lg}/v^{\rm rec}_{\rm lg,\parallel}(\beta_0=f) $
measured  in the 53 mock catalogs. These histograms are analogous to that 
obtained in Section~\ref{sec:dmrec} (black solid histogram) but 
refer to mock galaxies in real (blue, dot-dashed curve) and
and redshift (red dashed) space rather than to dark matter particles.


In real space these histograms represent the distribution of $\beta/f$, according to  (\ref{eq:betapar}).
In redshift space the link between the recosntructed LG mostion and the value of $\beta$ is less straightforward since, in this case, 
peculiar velocities do not scale linearly with $\beta$, as we show explicitly in Appendix B. 
The mean values and the scatter of each histogram are listed in Table~\ref{tab:histo} 

\begin{table}
\begin{center}
\caption{\label{tab:histo} Mean values and standard deviations of the 
distributions for   $v^{\rm tru}_{\rm lg}/v^{\rm rec}_{\rm lg,\parallel}(\beta_0=f) $
computed  from the 53 mocks and shown in  Figure~\ref{fig:beta}.
 Col.1: Tracers. Col 2: Type of reconstruction (real or redshift space).
 Col 3: Mean value. Col 4: Standard deviation.
  }
\label{tab:error}
\begin{tabular}{|c|c|c|c|}
\hline \hline
Tracer   & Space & Average & Variance   \\ 
\hline 
Dark Matter   & Real & 0.93 & 0.08 \\  
Mock Galaxies  & Real  & 0.80 & 0.09 \\ 
Mock Galaxies  & Redshift & 0.72 &  0.12  \\ 
\hline
\end{tabular}
\end{center}
\end{table}

The three histograms in the plot are not expected to match for several reasons. First of all, as previously discussed, 
linear theory is not quite able to describe the LG motion. Not even using the full dark matter particles population in real space.
This explains why the peak of the solid curve is at $\sim 0.9$, rather than $1$.
Secondly, galaxy bias induces a systematic mismatch between the value $ \beta_{\rm dm}$ obtained from dark matter particles 
and the one obtained from mock galaxies in real space,  $\beta_{\rm r}$ that should be equal to the linear bias parameter of the sample
 $\beta_{\rm r}/\beta_{\rm dm} \equiv b$. Indeed,  we find that  the value of this ratio  ($ =1.28$)  is consistent with the 
value of the linear bias in the mocks ($1.2-1.35$) obtained
from the scatterplot  (\ref{fig:bias}) in \S\ref{sec:biasing}.
This is a remarkable results since it shows that  comparing gravity and velocity in a single LG-like region
can provide an unbiased, if noisy, estimate of $\beta$.

Additional errors caused by performing the reconstruction in redshift rather than real space are the origin of the 
mismatch between the red-dashed and the blue, dot-dashed histograms. Remarkably, the differences between 
the two curves are not large. This is very welcome as putting all the galaxies into real space is problematic; it is easier
 to leave the galaxies in redshift space.

Assuming that in the observations $\beta$ is determined as outlined above, we further 
ask how well the direction of the LG motion can be recovered. 
The individual $\beta$ values above  yield $v^{\rm rec}_{\rm lg,\parallel}=v^{\rm tru}_{\rm lg,\parallel}$ 
  and, therefore, 
the angle, $\theta$,  between $\vv^{\rm rec}_{\rm lg}$ and $\vv^{\rm tru}_{\rm lg}$ is 
\[ 
\theta=\tan^{-1} \frac{v^{\rm rec}_{\rm lg,\perp}}{v^{\rm tru}_{\rm lg,\parallel}} \; .
\]
This yields $<\theta^2>^{1/2}\approx 10^\circ$ for real as well as redshift space reconstruction.

\subsubsection{$\vv^{\rm rec}_{\rm  lg}$ versus $\vv^{\rm tru}_{\rm  lg}$ for a fixed $\beta$}


We now assess the goodness of the LG velocity reconstruction, in both real and redshift space, when the value of $\beta$ is given 
{\it a priori}.  We do this in three steps. First we set a convenient value of $\beta$. Then we reconstruct $\vv_{\rm lg}$ using all 
objects within $R_{\rm out}=250\hmpc$ in each of the 53 mocks. And finally we compare the result with the true LG motion.
We set  $\beta$ in correspondence of the mean values
of the histograms shown in  Figure~\ref{fig:beta}, i.e. $\beta_{\rm r}=0.33$ and  
$\beta_{\rm s}=0.25$ in real and redshift space, respectively. 
This guarantees that the reconstructed parallel LG velocities are evenly distributed around the true values.

\begin{figure} 
\includegraphics[width=.48\textwidth]{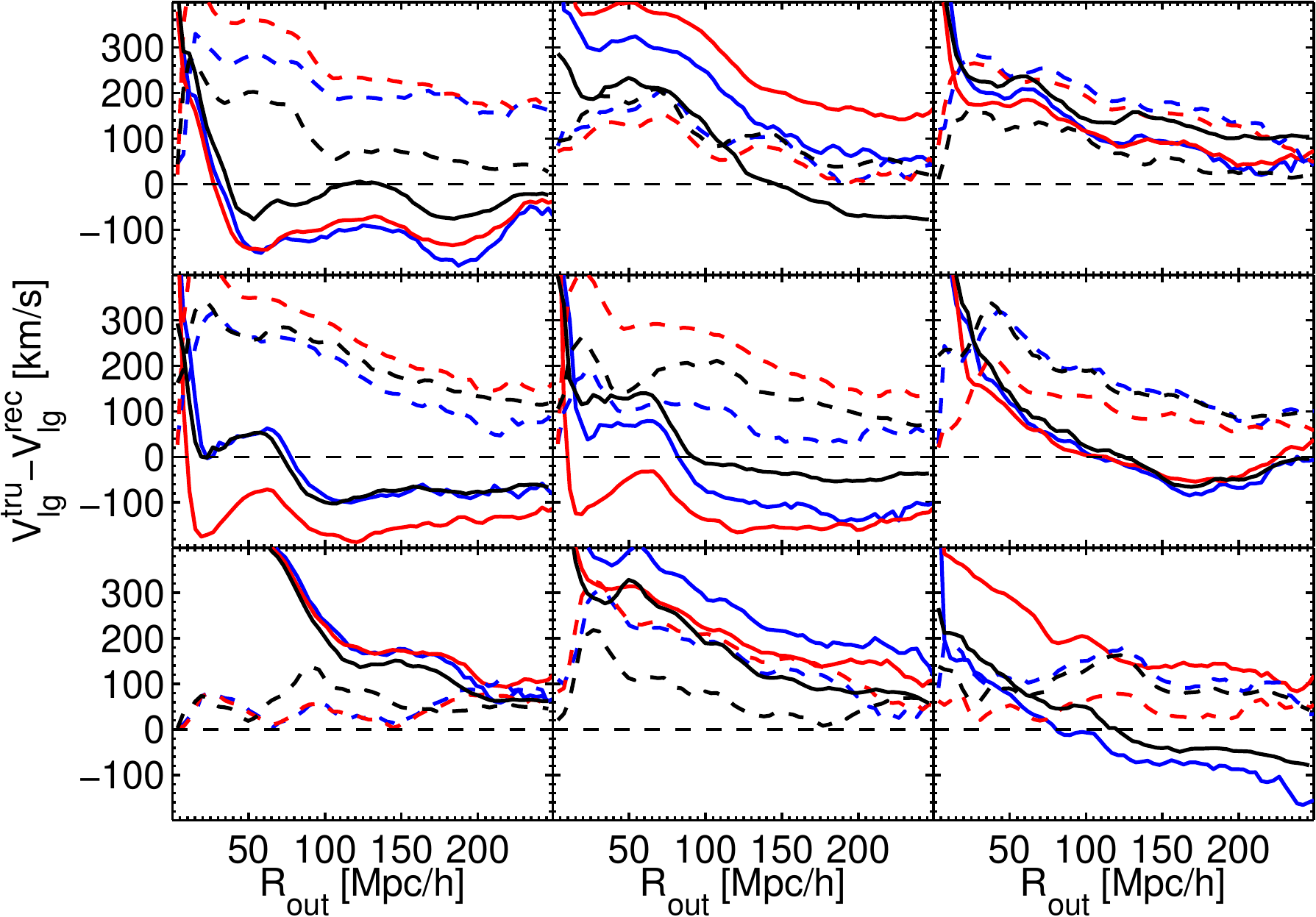}
\caption{ Residuals of  LG motions for the same 9 mocks in figure (\ref{fig:vdm})
 for   reconstructions from  the galaxy distribution in real (blue curves) and redshift (red) space. 
 For reference we also plot the black curves of figure (\ref{fig:vdm}) that represent the dark matter case.
 Parallel and perpendicular components of the residuals are shown as the solid and 
 and dashed  lines, respectively. }
 \label{fig:vmock}
\end{figure}

The results  are shown in Figure (\ref{fig:vmock}), which is the analogous of  Figure (\ref{fig:vdm}).
The 9 panels refer to the same mocks of that plot. Blue and red curves refer to reconstructions performed in
real and redshift space, respectively. The black curves are the same of Figure (\ref{fig:vdm}) and show 
the case of  dark matter reconstruction.
Moreover the residuals in redshift space are occasionally much smaller than in real space. This is an additional confirmation that there is no problem in performing the computation in redshift space \citep{ND94}.
Finally, we note that beyond  $150 \, \hmpc$ the curves become flat, indicating that most of the contribution to $\vv^{\rm rec}_{\rm lg}$   arises from large scale structure within that radius. 


\begin{figure} 
\vspace{0pt}
\includegraphics[width=0.45\textwidth]{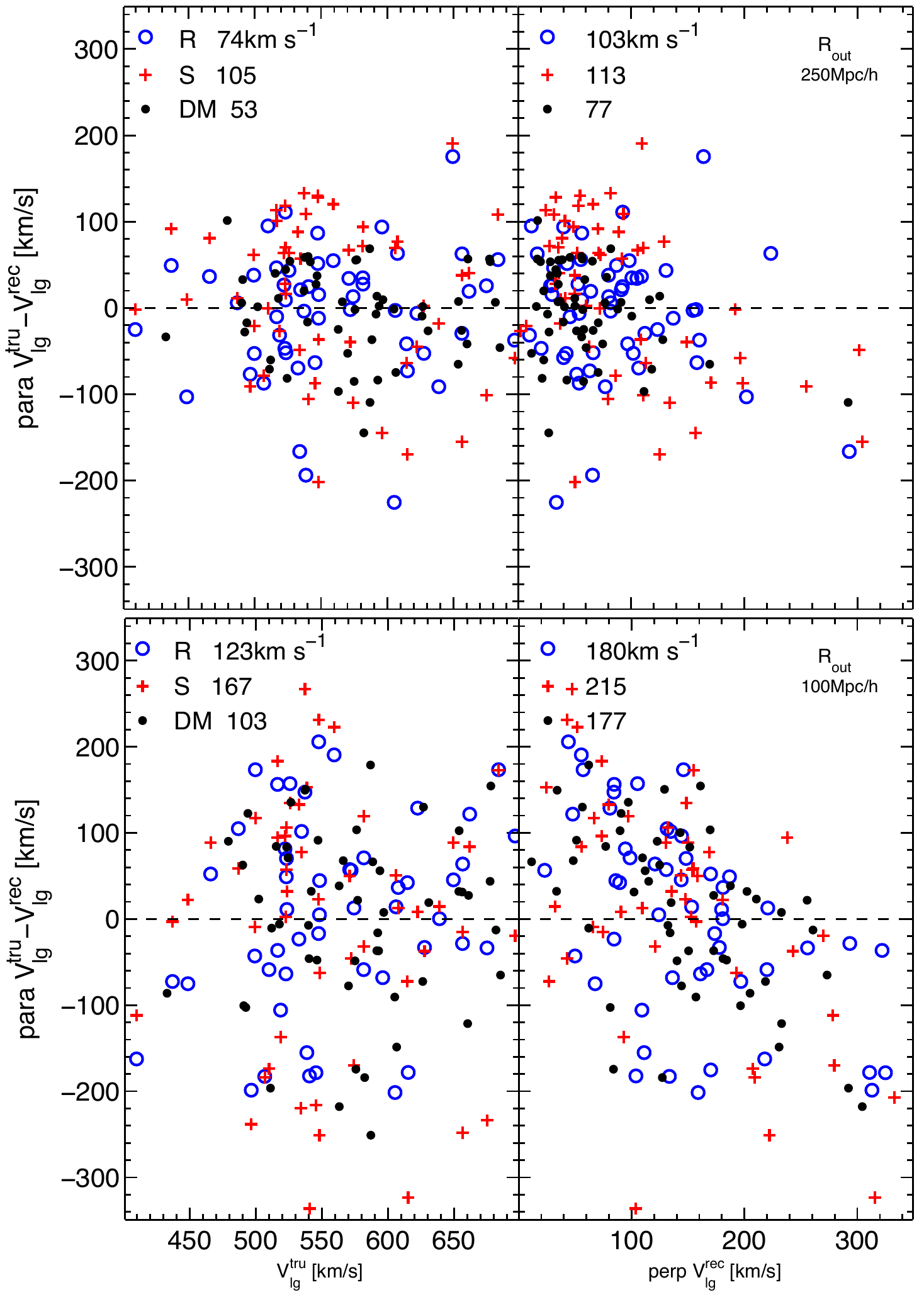}

\caption{A scatter plot showing  the velocity residual in the parallel and perpendicular directions  for all 53 mock catalogs. 
The blue dots are in real space and the red crosses are in  redshift space, while black dots show recovery from the full dark matter density field in real space. 
The  {\it rms} values of the  parallel and perpendicular residuals are listed in the left and right panels, respectively.
  Top and bottom panels correspond to velocity reconstruction with $R_{\rm out}=250\hmpc$
 and $R_{\rm out}=100\hmpc$, respectively.
The  {\it rms}  of the parallel and perpendicular residuals are indicated, respectively,  in the left and right panels. 
  }
\label{fig:Scatter_dvdv}
\end{figure}

Figure (\ref{fig:Scatter_dvdv}) provides an additional assessment of the goodness of the reconstruction. It shows
the parallel  velocity residuals vs.  the true velocity $\vv^{\rm tru}_{\rm lg}$ (left panels)  and vs. the perpendicular component of the 
reconstructed velocity,  $v^{\rm rec}_{\rm lg,\perp}$ (right panels) in each of the 53 mocks, 
The filled black dots refer to the case of dark matter particles reconstruction in real space.
Open blue dots and red crosses refer to reconstructions with mock galaxies in real and redshift space, respectively.
In the upper panels the reconstructed quantities are measured at 
$R_{\rm out}=250\hmpc$, to match the simulation size. This represents the ideal and rather unrealistic case of a deep, all-sky survey
with a selection function accurately estimated out to very large distances. The bottom panel, in which $R_{\rm out}=100\hmpc$,
represents a more realistic case in which, like in the 2MRS, the  
errors in the selection function are reasonably small out to $\sim 100\hmpc$ \citep{2012MNRAS.424..472B}.

The fact that in all plots the mean  of the parallel residuals is zero is just a consequence of the choice of  the  
$\beta$ value used in the reconstructions. Instead, the fact that the parallel residuals
are uncorrelated with $\vv^{\rm tru}_{\rm lg}$ but anti-correlated with $v^{\rm rec}_{\rm lg,\perp}$ is a genuine result.
The  anti-correlation implies some degeneracy between the error in the estimation of the direction 
and amplitude in the reconstruction of the LG motion and it should be kept in mind in realistic applications.
The accuracy of the LG velocity reconstruction is quantified by the  {\it rms} of the parallel residuals and is indicated in the plot.
As expected it is smallest when the reconstruction is performed in real space with dark matter particles and larger 
when one considers mock galaxies at their redshift space positions. 
Moreover, the scatter for $R_{\rm out}=100\hmpc$ is about twice as large as in the respective reconstructions for  $R_{\rm out}=250\hmpc$.
In the realistic case in which the reconstruction is performed in redshift space using all mock 2MRS galaxies
within $100\hmpc$ the  {\it rms } scatter is of the order of $272\kms$, a value that represents the typical error
on the estimated LG motion from currently available all-sky surveys.

The scatter plot in Figure (\ref{fig:Scatter_dvdv}) can help to investigate in detail the error budget of the LG velocity reconstruction.
Let us consider the case of $R_{\rm out}=250\hmpc$.
For LG velocity reconstructions with mock galaxies in real space the {\it rms } of the total residual, i.e. is the sum in quadrature of the {\it rms} values of the parallel and perpendicular residuals, is  $127\kms$. This is higher that the {\it rms} value of $93\kms$ obtained using dark matter particles.
Part of the difference is accounted for by shot noise which provide an additional contribution of  $\sim 70\kms$,
as shown in S\ref{sec:sn}. We attribute the remaining, small discrepancy,  to the stochastic nature of the biasing relation seen in figure (\ref{fig:bias}).
In redshift space the total scatter of the residuals is about $20\%$ higher than in real space.
This additional uncertainty must be due to non-linear effects which leak  differently in real and redshift space
and to the multi-valued nature of the real-to-redshift space mapping in regions of high density.

\section{The ``Kaiser rocket" effect}
\label{sec:kaiser}
\begin{figure} 
\includegraphics[width=.45\textwidth]{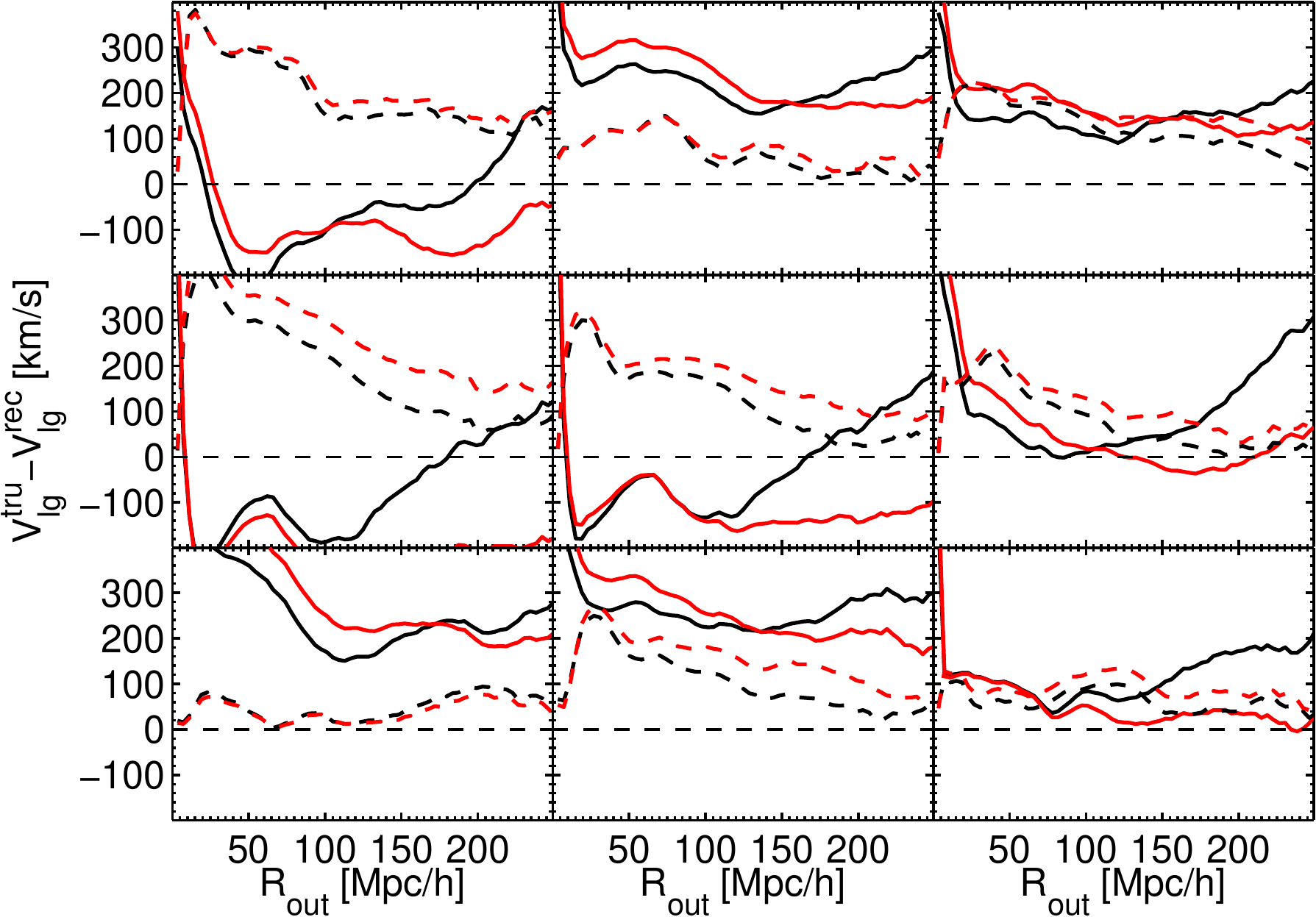}
\caption{Illustration of the rocket effect in the recovery of LG motion.
Black curves are residuals from redshift space reconstructions with galaxies weighted by the selection function evaluated at the 
redshift space coordinate rather the actual distance. Red curves are  result of correcting for the rocket effect, as described in the text,  in the recovery of the LG motion. Solid and dashed curves correspond to parallel and perpendicular residuals.     }
\label{fig:LGrocket030}
\end{figure}

Redshift surveys are characterized by different types of selection effects that may depend on the 
intrinsic properties of the objects and the distances. These effects are  quantified 
by a selection function, $\varphi$. Let us focus on the distance dependence and consider the case of
a flux limited survey, like 2MRS.  Galaxies must be weighted by the inverse of the selection 
function to compensate for the 
unobserved faint galaxies with fluxes falling below the detection limit.
The weight assigned to a galaxy should be proportional to $\varphi^{-1}(r)$ evaluated at the unknown  distance to the 
galaxy, whereas in practice the selection function is measured at the redshift of the galaxies.
 All tests used above have indeed  used $\varphi^{-1}(r)$.
The use of  $\varphi^{-1}(s)$, the selection function evaluated at the redshift of the galaxy,
 leads to systematic biases in the reconstruction of the LG motion and bulk flows: the so-called
  ``Kaiser rocket" effect  \citep{K87}.  
  In order to demonstrate the importance of the effect we have recovered the LG motion with galaxies weighted by  $\varphi^{-1}(s)$  rather than $\varphi^{-1}(r)$.  The corresponding residuals
 are  shown as black curves  in figure (\ref{fig:LGrocket030}).   
In a survey like 2MRS the ``Kaiser rocket" effect remains   tamed  for  $R_{\rm out}\ltsim 100\hmpc$, but increases substantially  at larger distances, overshooting to large values as  $R_{\rm out}$ approaches $250\hmpc$. 
  This behavior explains why we have presented results for $R_{\rm out}=100\hmpc$ in figure (\ref{fig:Scatter_dvdv}) as the appropriate value for 2MRS-like surveys.

There are several ways one could attempt a correction for this effect. 
To linear order, this introduces a correction term to the general linear theory relation in redshift space \citep{ND94}, which could be solved 
directly using standard numerical method. This will yield an estimate of the velocity field, $\vv$, of galaxies in the LG frame, from which the 
reflex dipole at large distances could be computed and identified with the negative of  LG motion with respect to the CMB, as explained in the Appendix. 
Another strategy is to adopt an iterative approach.  At the iteration step $i$ the selection function  is  computed  at $\tilde r=s- \tilde r\hvs\cdot \vv^{i-1}/H_0$ where $\vv^{i-1}$ is the reflex dipole motion of a shell obtained from the previous iteration, starting  with zero at the first iteration.
In these iterations,  the reflex dipole approximates the velocity field of galaxies at the same redshift. This is rigorously correct to linear order where the Kaiser effect does not mix different multipoles of the velocity field, leaving the dipole as the only relevant component for the recovery of the LG motion. 
This iterative scheme allows the use of the new relation  (equation \ref{eq:vlgs}) derived here.

In practice, these iterations are time consuming  and to assess the impact of the effect in this Section
we simply use 
$\varphi(\tilde r)$ with $\tilde r$ given from the reflex dipole (equation \ref{eq:vlinsd}) recovered from linear theory
using the correct selection function $\varphi(r)$ evaluated  at the actual 
galaxy distance, $r$.
The residuals in the reconstruction including this correction are plotted  as the red curves in figure (\ref{fig:LGrocket030}). 
The correction manages to suppress the overshooting at large distances and brings the reconstruction  closer to corresponding 
curves in figure (\ref{fig:vmock}). But the total {\it rms } scatter is still significant- $200\kms$ at $R_{\rm out}=250\hmpc$. 
We emphasize that this is an oversimplified test of the correction to the Kaiser effect. For real catalogs, the determination of the selection function 
invokes an assessment of galaxy evolution that might be significant even within $R\gtsim 100\hmpc$. 
Further, the short cut we have taken to derive the distance  $\tilde r$ assumes that the final outcome of the iteration procedure 
is as accurate as using the reconstruction with $\phi(r)$.  Had this been true, the red curves in figure  (\ref{fig:LGrocket030}) would have coincided 
with the corresponding curves in figure  (\ref{fig:vmock}), which is not the case.  Therefore, the correction to Kaiser in realistic application is 
far more uncertain. 

\section{Discussion}
\label{sec:disc}

In this paper we have investigated the various error sources  in the determination of the LG motion.
\begin{itemize}

\item It is a mistake to do the analysis in the CMB frame because this frame is only gradually reached. The LG frame gives a better estimate of the distance to the nearby mass fluctuations, as seen in the substantial deviations in the residual vector differences for $R_{out} < 70\hmpc$  (e.g.  figure (\ref{fig:vdm})).

\item By far the main source of error is related to the limited depth of galaxy surveys, i.e. cosmic variance.  
The error in the LG motion estimated from the full dark matter density field in 
real space and within a radius of  $250\hmpc$ is  $\sim 90\kms$.  
However, the dilute sampling and flux-limited nature of
most avalable datasets makes them significantly shallower than this.
 For an all sky survey like the 2MRS,
the contribution to the LG motion can be assessed reliably only within 
 $\sim 100\hmpc$. 
 At this  depth, the error in the predicted LG motion is $\sim 200\kms$.

\item The velocity residuals measured in the perpendicular and parallel directions allow us to estimate the {\it rms} angle of the 
LG velocity from the full sky data. We find that errors of $\approx 10^\circ$ are inevitable, and are not sensitive to whether the analysis is done 
in  redshift space or real space.

 \item 
Another source of uncertainty are the  errors in the dynamical reconstruction.
To assess their contribution to the total error budget  we have estimated 
the LG motion from the full dark matter out to 
the largest possible outer  radius, i.e. $250\hmpc$. 
Then, we have filtered out the contribution from large scale structure beyond this radius
by measuring the LG motion with respect to the bulk flow of the sphere.
Hence the resultant errors are entirely due to dynamical inaccuracies of linear theory. 
The corresponding $1\sigma$ error, $\sim 90\kms$, is substantially  smaller that the typical error  in linear reconstruction 
of the peculiar velocity of a generic observer in the Universe.
The reason for this is the strict criteria we 
have applied in  selecting the "LG observer" in the mock catalogs, aimed at  matching   the quietness and moderate density  environment of the 
observed LG.   Removing these selection criteria boosts the  error to $\gtsim 300 \kms$, consistent with previous studies \citep{NB00}.

Nonlinear dynamical reconstruction methods 
 \citep[e.g.][]{sh95,cc97,f02,NB00,BEN02} can potentially reduce the dynamical error.
 However, because the particular environment of the LG,  errors due to linear reconstruction
 are subdominant compared to the total error budget discussed above.  
 
 \item
The shot noise originating from the 
sampling of the mass density field by a finite number of tracers, forcing the use of  
the selection function of the sample $\varphi$ to deal with the magnitude limit,  
is another error source.
In the case of the 2MRS galaxies brighter than K$_{\rm s}=11.75$,  the error amplitude is 
comparable to that of the dynamical error.
The {\it rms} of the combined  dynamical and shot noise scatter   
is close to the error  estimated directly from the scatter of the 
LG motions reconstructed from our mock galaxy catalogs, leaving 
little room for any substantial, additional source of error.

\item
Indeed, we have identified the remaining error source with galaxy biasing.
Galaxies in our mocks follow a  biasing relation that is close to linear in regions with positive densities
but it significantly deviates from linearity in voids where
galaxies are less abundant than expected from an extrapolation of the linear bias.
Moreover, the biasing relation is non-deterministic. Its scatter is driven by Poisson noise 
everywhere but in high density regions, where the intrinsic scatter in the biasing relation
is dominant.
Nonetheless, uncertainties due to deviations from the assumptions of linear biasing are of relative insignificance as 
illustrated by the comparison of  the real space reconstructions from 2MRS-like mocks generated from dark matter particles and 
 the mock galaxy catalogs.

 \item
 Reconstructing velocities from redshift space data is fundamentally a more challenging problem than 
 in real space. Effects like multi-valued zones (tracers with distinct distances along the same line of sight, but with very similar redshifts)
 and fingers of god (spreading of galaxies in virialized regions along the line of sight) affect the reconstruction on scales larger than the traditional non-linear scale.  In order to mitigate the effect of the fingers of god, the redshifts of the mock galaxies have been computed with 
 peculiar velocities smoothed on a scale $1\hmpc$ scale.   
The $1\sigma$ errors in the LG motion reconstructed in
redshift space reconstruction is  $\sim 20\%$ larger than in real space. 

\item 
We have also assessed the "Kaiser rocket" effect and demonstrated that it can be  partially corrected
if the selection function is well constrained by observations. Correction is easy in the case of  mock catalogs
but more challenging in real datasets where the effect of galaxy evolution cannot be ignored, even within
 the local volume encompassed by
the 2MRS  \citep{2012MNRAS.424..472B}.
A distinct signature of the "Kaiser rocket" effect is the overshooting of the reconstructed LG motion 
at large radii, ($\gtsim 100\hmpc$).
 We are not aware of any reconstruction of galaxy dipoles from real datasets taking into account this spurious growth.

\end{itemize}

Despite the general consensus on its origin, the identification of 
the actual gravitational sources  responsible for the LG motion is still a matter of debate.
Gravity is a long range force and it may prove futile to try to identity 
specific sources for the LG motion. 
A more useful description is  in terms of  
  ``dipole convergence scale", i.e. the physical distance which encompasses 
the matter fluctuations responsible  for generating  most of the LG motion.
In the standard $\Lambda$CDM model, the dominant contribution to the LG motion 
is expected to arise from mass fluctuations with a distance up to $\sim 200-300\hmpc$ from the LG \citep{bilicki11}. 
We have 
seen that convergence in the mocks is achieved at a  depth of $\approx 250\hmpc$, consistent with the theoretical expectation.
The convergence is gradual, indicating that the cumulative effect of the large scale mass density field must be invoked to account for the LG motion,  
   
The  $150-200\kms$ accuracy  should be regarded as a conservative estimate for the expected
error in the estimates of the LG velocity.
Indeed,  comparisons between the observed LG motion and the prediction from various redshift surveys 
have yielded consistency up to the level we see here in the mocks. 
We, therefore, conclude that the standard picture for  the formation of large scale structure is fully consistent with  
current observational data of the large scale motions.

\section{acknowledgment}
Special thanks are due to Gerard Lemson for providing us with data from the German Astrophysical Observatory (GAVO).
This research was supported by the I-CORE Program of the Planning and Budgeting Committee,
THE ISRAEL SCIENCE FOUNDATION (grants No. 1829/12 and No. 203/09), the
German-Israeli Foundation for Research and Development, and the Asher Space Research
Institute.
EB  acknowledges the financial support provided by
MIUR PRIN 2011 'The dark Universe and the cosmic evolution of baryons: from current surveys to Euclid'
and  Agenzia Spaziale Italiana for financial support from the agreement ASI/INAF/I/023/12/0. 
AN and EB Thanks the Department of Astronomy at the University of Cape Town for hospitality. MD acknowledges  funding from CAASTRO/Swinburne  where he completed this work.

\bibliography{LGM.15}

\begin{appendix}
\section{Reconstructing LG motion}
\label{sec:appendixa}
 \subsection{Relations for continuous fields}
The LG motion is a special case of the bulk flow. 
Thus we first derive   a linear relation between the  bulk flow and the density contrast $\delta(\vr)=\rho(\vr)/\bar \rho-1$.
The  bulk flow of a sphere of radius $R$ centered at the origin 
\begin{equation}
\label{eq:Bvol}
\vB=\frac{1}{V}\int_{_{r<R}}\dd r^3 \vv(\vr) \; .
\end{equation}
where $V=4\pi/3R^3$.

We note the mathematical identity\footnote{This is a particular form of the 
Gauss (or Green) Theorem  $\int_V \grad  \cdot \vA \dd^3 r=\int \vA \cdot \dd \vS$ when one consider a scalar 
field $\vA=A$ and  $\grad  \cdot = \grad$}  $\int_V \grad g \dd^3 r=\int_S g \dd \vS$ where $g$ is a scalar function,
the vector $\dd \vS$   is an element of the surface enclosing the volume $V$. 
Applying  this identity with $g=-\phi$ and noting the definition of $\vB $  in (\ref{eq:Bvol}), yields
\begin{eqnarray}
\nonumber \vB&=&-\frac{1}{V}\int_S \phi \dd \vS\\
 & =& -\frac{1}{V}\int \phi(|\vs|=r, \hvs) \hvs r^2 \dd \Omega\; ,
\label{eq:surf}
\end{eqnarray}
where the second line is valid for  a spherical surface  of radius $R$ centered on the origin. 
We substitute  $\phi$ in terms of spherical harmonics expansion over $\phi_{lm}$ from (\ref{eq:vlins}).
Thanks to  the orthogonality condition $\int Y_{lm} Y^*_{l'm'}\dd \Omega=\delta^K_{ll'}\delta^K_{mm'}$, 
only $l=1$ contributes to the integral, with the net result, 
\begin{eqnarray}
\nonumber \vB&=& -\frac{H_0\beta/(1+\beta)}{2\tl+1} V^{-1}
\left[     r^{\tl+2}\int_0^r \frac{\vD(u)}{u^{\tl-1}}\dd u
 \right.  \\
 &-& \left.   r^{1-\tl} \int_0^r \vD(u)u^{\tl+2}\dd u      \right]\; ,
 \end{eqnarray}
 where $\Delta_z=\sqrt{4\pi/3}\delta^{\rm s}_{_{10}}$,   $\Delta_y=i \sqrt{2\pi/3}(\delta^{\rm s}_{_{11}}+\delta^{\rm s}_{_{1\, -1}})$, 
 $\Delta_x=- \sqrt{2\pi/3}(\delta^{\rm s}_{_{11}}-\delta^{\rm s}_{_{1\, -1}})$ and $(1+\beta)\tl(\tl+1)-2=0$ since $l=1$.
 In the derivation we have also  used the relations
 $\hvz \cdot \hvr=\cos\theta=\sqrt{4\pi/3}Y_{_{10}}$, $\hvy \cdot \hvr=\sin\theta \sin\varphi= i \sqrt{2\pi/3}(Y_{_{11}}+Y_{_{1\, -1}})$ and 
 $\hvx \cdot \hvr=\sin\theta\cos\varphi=- \sqrt{2\pi/3}(Y_{_{11}}-Y_{_{1\, -1}})$.  
 
 The bulk motion of a thin shell, i.e. the reflex dipole motion of the shell, 
 is defined as 
 \begin{equation}
 \vv^{\rm shell}(r)=\frac{1}{4\pi}\int \vv(\vr)\dd \Omega \; .
 \end{equation}
 Thus 
 \begin{equation}
 \label{eq:shell}
   \vv^{\rm shell} =\frac{1}{3R^{2}} \frac{\partial [R^3 \vB(R)]}{\partial R}
 \end{equation}
For completeness, we also note that an application of the mathematical identity above to compute the mean motion of a shell of radius 
$r$ and thickness $\delta r\rightarrow 0$, yields 
\begin{equation}
  \vv^{\rm shell}=  -\frac{1}{4\pi}\int \left[\frac{\dd \phi}{\dd r} +2\frac{\phi}{r}\right]\dd \Omega \; .
 \end{equation}

    The LG velocity, $\vv_{\rm lg}$,  is the  negative of  the reflex dipole of the  shell at  infinity so that,
 \begin{equation}
\vv_{\rm lg}=-\lim_{r\rightarrow \infty} \vv^{\rm shell}(r)\; .\
\end{equation}

Using (\ref{eq:vlinsd}) and (\ref{eq:shell}) we get 
\begin{eqnarray}
\nonumber \vv^{\rm shell}&=& -\frac{H_0\beta/(1+\beta)}{2\tl+1} (3V)^{-1}
\left[   (\tl+2)  r^{\tl+2}\int_0^r \frac{\vD(u)}{u^{\tl-1}}\dd u
 \right.  \\
 &-& \left.  (1-\tl) r^{1-\tl} \int_0^r \vD(u)u^{\tl+2}\dd u      \right]\; .
 \label{eq:vlinsd}
 \end{eqnarray}

 \subsection{Relations for a discrete sampling}
 We now modify the above relations for the LG motion and the bulk flow to the case of a discrete sampling of density field in redshift space 
by   a discrete distribution of $N$ tracers, $i=1\cdots N$, with mean number density $\bar n$.
This modification will allow an  application of linear theory reconstruction directly on the distribution 
of galaxies in redshift space, rather then employ a smoothing procedure in order to use the relations above. 
In the discrete representation, the density field is approximated as
 \begin{equation}
 \delta(\vs)={\bar n}^{-1}\sum_{i=1}^N  s_i^{-2} \delta^D(\vs-\vs_i) \; 
 \end{equation}
 where $\delta^D$ is Dirac's delta function and $\vs_i$ are the redshift coordinates of the tracers. 
 If the tracers are galaxies in a flux limited survey, then the 
 contribution of each point should be weighted by a selection function to account for the loss of fainter galaxies at 
 larger distances. For brevity of notation, at this stage we assume a volume limited survey so that
 all tracers are equally weighted. 
The spherical harmonics expansion  is
\begin{equation}
 \delta^{\rm s}_{lm}={\bar n}^{-1}\sum_{i=1}^N  s_i^{-2} \delta^D(s-s_i) Y_{lm}(\hvs_i)\; ,
 \end{equation}
giving 
 $\vD={\bar n}^{-1}\sum_i s_i^{-2} \delta^D(s-s_i) \hvs_i$.  The expression 
(\ref{eq:vlinsd}) for the bulk flow becomes
\begin{equation}
 \vB(r) = \frac{-r^{\tl+2}}{(1+\beta)(2\tl+1)}\frac{\beta H_0}{\bar  n V}\sum_{s_i<r}\frac{\vs_i}{s_i^{\tl+2}}
 +\frac{r^{1-\tl}}{(1+\beta)(2\tl+1)}\frac{\beta H_0}{\bar n V}\sum_{s_i<r}s_i^{\tl-1}\vs_i\; .
\label{eq:bdenss} 
\end{equation}
Another form of this relation which should be more appropriate for numerical applications is
\begin{equation}
 \vB(r) = \frac{-\beta H_0}{(1+\beta)(2\tl+1)}\left[r^{\tl+2} \left<\frac{\vs_i}{s_i^{\tl+2}}\right>_r
-r^{1-\tl}\left<s_i^{\tl-1}\vs_i\right>_r \right]\; .
\label{eq:bdensst} 
\end{equation}
 where   $<X_i>_r=\sum_{s_i<r} X_i/(\bar n V)  $ and is approximated numerically 
 by averaging over particles, i.e. $\bar n V$ is approximated as the number of particles within $r$.
 This form avoids potential problems due to the divergent behavior  of $r^{\tl+2}/V\sim r^{\tl -1}$ as $r \rightarrow 0$.
 and
\begin{equation}
\vv^{\rm shell}(r) = \frac{-r^{\tl-1}}{(1+\beta)}\frac{\beta}{4\pi\bar n}\sum_{s_i<r}\frac{\vs_i}{s_i^{\tl+2}}
+ \frac{1-\tl}{2\tl+1}\frac{r^{-(2+\tl)}}{(1+\beta)}\frac{\beta}{4\pi\bar n }\sum_{s_i<r}s_i^{\tl-1}\vs_i\; .
\label{eq:bdenshell} 
\end{equation}


\begin{equation}
 \vv^{\rm shell}(r) = \frac{-\beta H_0}{(1+\beta)(2\tl+1)}\left[\left(\frac{2\tl+1}{3}\right) r^{\tl+2} \left<\frac{\vs_i}{s_i^{\tl+2}}\right>_r
-\left(\frac{1-\tl}{3}\right)r^{1-\tl}\left<s_i^{\tl-1}\vs_i\right>_r \right]\; .
\end{equation}

In the  relations (\ref{eq:vlinsd}) and (\ref{eq:bdenss})
it is understood that the sphere of radius $r $ is centered at the origin, $\vr=0$,  of the coordinate system in real space. 
Since we work in the LG frame, $\vr=0$  corresponds to $\vs=0$, hence the sphere is also  centered at the origin in redshift space. 
We note that the contribution of the second term on the r.h.s of (\ref{eq:bdenss}) becomes increasingly small as $R\rightarrow 0$. 
This is sustained analytically by the form of this term which gives more weight to larger distances where homogeneity 
is more pronounced and is corroborated by the analysis of the mock catalogs in Section~\ref{sec:dmrec}.
As a consequence the LG velocity  can be expressed as
\begin{equation}
\vv_{\rm lg}= \frac{(2\tl +1)R_{\rm out}^{\tl-1}}{(1+\beta)(2\tl+1)}\frac{\beta H_0}{4\pi\bar n}\sum_{R_{\rm lg}<s_i<R_{\rm out}}\frac{\vs_i}{s_i^{\tl+2}}
 -\frac{(1 -\tl)r^{1-\tl}}{(1+\beta)(2\tl+1)}\frac{\beta H_0}{3\bar n V}\sum_{R_{\rm lg}<s_i<R_{\rm out}}s_i^{\tl-1}\vs_i\; .
\label{eq:vlgsapp} 
\end{equation}
which also appears as equation (\ref{eq:vlgs}) in the main body of the paper.
The sum extend over all objects between $R_{\rm lg}$,  the radius assigned to the LG, and 
a maximum distance $R_{\rm out}$.
Note that there is no lower cutoff at $R_{\rm lg}$ in the bulk flow expression (\ref{eq:bdenss}) 
simply because the region within $R_{\rm lg}$ should be counted as part of the sphere 
$R$ for which the bulk flow is computed. The fact that $R_{\rm lg} \ll R$ 
makes the cutoff in (\ref{eq:bdenss})  insignificant.
The maximum distance $R_{\rm out}$ should be 
assumed to approach infinity in the ideal case where galaxies are observed over all space. 
However, in realistic redshift surveys, the number density of galaxies decreases with redshift 
 due to magnitude cuts, making the noise increase dramatically at larger redshifts. 

The real space counterparts of the relations (\ref{eq:bdenss}) and (\ref{eq:vlgsapp}) are obtained by 
setting $\tl=l=1$ and replacing $f$ with  $\beta/(1+\beta)$. This yields 
 \begin{equation}
\vB=\frac{-H_0\beta}{4\pi \bar n} \sum_{r_i<R}\frac{\vr_i}{r_i^3}+\frac{H_0\beta}{3\bar n V}\sum_{r_i<R} \vr_i\; ,
\label{eq:bdens}
\end{equation}
and
\begin{equation}
\vv_{\rm lg}=\frac{H_0\beta}{4\pi \bar n} \sum_{R_{\rm out}>r_i>R_{\rm lg}}\frac{\vr_i}{r_i^3}\; .
\label{eq:vlgapp}
\end{equation}
Since $\bar n V\approx  N$, the total number of objects in the sphere, we see that the second term 
in the r.h.s. of the bulk flow expression  (\ref{eq:bdens}) is proportional to the center of mass coordinate,  
$\sum \vr_i/N $, as already noticed by  \cite{vittorio}. 


\section{The dependence on $\beta$ in reconstruction from redshift space data}
\label{sec:scaling}

In linear theory,  the recovered peculiar  velocity field in real space
is linearly proportional to  $\beta$. It can be shown that nonlinear dynamics 
preserve this proportionality  to a very good approximation \citep{1993ApJ...405..449G,NC98}.
In redshift space, the non-isotropic enhancement of the density by  the radial peculiar velocities
introduces a non-trivial  dependence on $\beta$. This is evident 
from equation (\ref{eq:vlgs}) in which deviations from linearity
arise from the explicit dependence of $\tl$ on $\beta$. 
We characterize this dependence by a function $F(\beta)=|\vv_{\rm lg}(\beta)|/|\vv_{\rm lg}(\beta=\beta_0)|$ 
where $\beta_0$ is a fixed reference value.
 
\begin{figure} 
\centering
\includegraphics[width=0.45\textwidth]{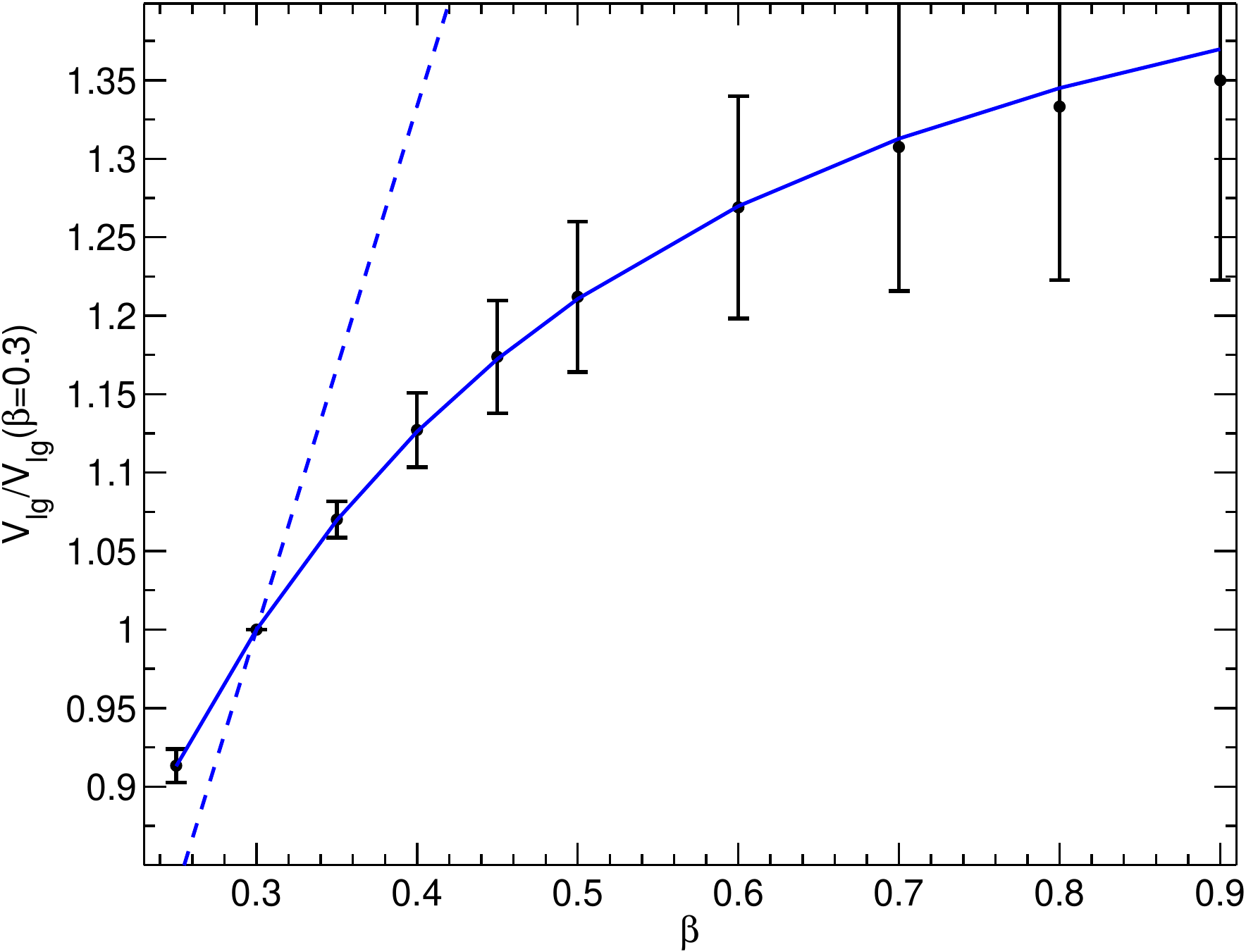}
\vspace{0pt}
\caption{The dependence of the redshift space reconstruction of  $\vv_{\rm lg}$ on $\beta$.  
 The black dots are the average, from all 53 mock,  of $|\vv_{\rm lg} |$ scaled by its value
at $\beta=0.3$ and the error-bars are  the corresponding {\it rms} scatter.
The blue  solid curve is   a fit, ${F_0 \beta}/{(1+7\beta^{3/2})^{2/3}}$, to the  the black dots. 
The dashed line is the dependence of real space reconstruction on $\beta$, i.e. $(\beta/0.3)$ .
}
\label{fig:beta_scale}
\end{figure}

Equation (\ref{eq:vlgs})  implies that $F$ depends also on
the actual distribution of tracers. Hence,  there is no  universal form for  $F(\beta)$ which is valid for any distribution.
Here we focus on the 2MRS and explore the dependence on $\beta$ from the  mocks. We proceed as follows.
For  each mock we perform the sum  in equation (\ref{eq:vlgs})
over all galaxies within $R_{\rm out}=250\hmpc$, for several values of $\beta$.
For each $\beta$,  we then compute  the mean and standard deviation of $ F(\beta)=|\vv_{\rm lg}(\beta)|/|\vv_{\rm lg}(\beta=0.3)$
from all the mocks. 
The black dots in figure (\ref{fig:beta_scale}) showing  the mean $  <F(\beta)>$,
significantly deviates from the linear scaling (dashed line) appropriate for real space reconstruction.
The standard deviation is represented by  the error-bars and it reflects the scatter in $F(\beta)$ due to variations of the 
galaxy distributions among the mocks. 
The blue, solid curve represents a polynomial fit to the black dots,   
\begin{equation}
\label{eq:fscale}
<F>\approx \frac{F_0 \beta}{(1+7\beta^{3/2})^{2/3}} \; ,
 \end{equation} 
 where $F_0$ is a constant such that $F(\beta=0.3)=1$.


\end{appendix}

\end{document}